\documentclass[reprint,prl,aps,showpacs,floats]{revtex4-1}

\usepackage{graphics}
\usepackage{amsmath}
\usepackage{amsfonts}
\usepackage{amssymb}
\usepackage{amsthm}
\usepackage{mathbbol}

\newcommand{\be}{\begin{equation}}
\newcommand{\ee}{\end{equation}}
\newcommand{\bea}{\begin{eqnarray}}
\newcommand{\eea}{\end{eqnarray}}
\newcommand{\ba}{\begin{array}}
\newcommand{\ea}{\end{array}}

\newcommand{\Del}{\ensuremath{\Delta}}

\newcommand{\noi}{\noindent}

\newcommand{\ket}[1]{\ensuremath{| #1 \rangle}}
\newcommand{\bra}[1]{\ensuremath{\langle #1 |}}
\newcommand{\ra}{\rangle}
\newcommand{\la}{\langle}

\begin{document}

\title{\bf Detecting entanglement of two electron spin qubits with witness operators}

\author{A. Borras and M. Blaauboer}

\affiliation{Kavli Institute of Nanoscience, Delft University of Technology,
Lorentzweg 1, 2628 CJ Delft, The Netherlands}

\date{\today}

\begin{abstract}
We propose a scheme for detecting entanglement between two electron spin qubits in a double
quantum dot using an entanglement witness operator. We first calculate the optimal configuration
of the two electron spins, defined as the position in the energy level spectrum where, 
averaged over the nuclear spin distribution, 1) the probability to have two separated electrons, and 
2) the degree of entanglement of the quantum state quantified by the concurrence are both large. Using a density matrix approach, we then calculate the evolution of the expectation value of the witness operator for the two-spin singlet state, taking into account the effect of 
decoherence due to quantum charge fluctuations modeled as a boson bath. We find that, for large interdot coupling, it is possible to 
obtain a highly entangled and robust ground state.
\end{abstract}

\pacs{03.67.Bg,03.67.Mn,73.21.La}

\maketitle

\noi 
Entanglement - non-classical correlations between quantum-mechanical particles - has for a long time been a theoretically predicted concept~\cite{epr35} without experimental proof. The first experimental observation of entanglement was reported for photon pairs in 1972~\cite{free72}, and since then also entanglement between protons~\cite{lame76}, kaons~\cite{bram99}, trapped ions~\cite{rowe01}, individual neutrons~\cite{hase03}, between an atom and a photon~\cite{moeh04}, and between superconducting qubits~\cite{stef06} has been demonstrated.
An important goal in present-day solid-state quantum physics is to generate and detect (prove) entanglement between {\it individual electrons}. The motivation behind this quest comes both from the fact that entanglement between electrons in a solid-state structure has so far not yet been demonstrated and from the recent
experimental progress in the field of quantum information processing in these systems~\cite{hans07}, which has, among others, led to
experimental realization of single- and two-qubit manipulations of electron spin qubits in quantum dots~\cite{pett05,nowa07} and coherent control of spins in diamond~\cite{hans08}.

In this Letter we propose a scheme for detecting entanglement in the former system, namely entanglement 
of two electron spin qubits in a double quantum dot. Many aspects of this quantum system, such as hyperfine 
coupling to the nuclear 
spins~\cite{kopp05,cois05,bluh10}, spin blockade~\cite{ono04,jour06}, and effects of applying a slanting magnetic field~\cite{toku06} are currently active topics of research. Our proposal consists of preparing and tuning the two-spin system such that the ground state 
contains a large entangled component and measuring a so-called entanglement witness operator~\cite{terh02} 
to demonstrate the presence of this entanglement. Entanglement witnesses are hermitian operators that are designed
to detect a specific entangled state. Their
expectation value is positive for all separable quantum states (in the class of states considered) and negative 
for at least one entangled state, usually the state the experiment aims to create. Entanglement witnesses have been used to detect entangled states in trapped ions systems or entangled photons states~\cite{guhn09}.

Theoretical proposals for detecting entanglement between individual electrons using witness operators are scarce.
Ref.~\cite{faor06} presents a proposal for implementing witness operators to detect electron-hole entanglement
in multiterminal conductors in the presence of noise due to random accumulated phases. In a previous work~\cite{blaa05},
we have proposed a turnstile mechanism as a suitable set-up for demonstrating entanglement between two electron spins
in a double quantum dot, assuming phenomenological decoherence times for spin relaxation ($T_1$-time) and
decoherence ($T_2$-time). In the present paper we focus on the dynamic evolution of the entanglement witness 
operator under the influence of decoherence due to quantum charge fluctuations, which are caused by changing gate voltages to control the system. Starting from the two-electron Hamiltonian,
and assuming a large external magnetic field, we first calculate the energy levels and eigenstates in the
three-dimensional Hilbert space spanned by the $|S(0,2)\ra$, $|S(1,1)\ra$ and $|T_0(1,1)\ra$ eigenstates, where 
$|S(n,m)\ra$ ($|T_0(n,m)\ra$) represents
the singlet (triplet) state with $n$ electrons in the left and $m$ electrons in the right dot. We then average over the nuclear field 
components and calculate the average probability of having one electron in each dot $\la P_{11} \ra_{\rm nucl}$ and the concurrence $\la C\ra_{\rm nucl}$ of the entangled component of the ground state. This allows us to determine the "optimal" configuration in the level diagram
in which both of the latter quantities are large. We then investigate the effects of decoherence of the entangled 
quantum state due to quantum charge fluctuations, modeled as a boson bath. Using the Born-Markov approximation, 
we calculate the reduced density matrix $\rho(t)$ of the system and the evolution of the expectation value of 
the witness operator Tr$(W\rho(t))$. We find the optimal values in our parameter space that provide a highly entangled ground state 
which is also robust under the action of charge fluctuations.

{\it Model. -} We consider a charge configuration of the double quantum dot where each electron can be located in a different quantum dot (1,1), or both of them in the right dot (0,2). In the (1,1) configuration the four accessible states are the singlet $|S(1,1)\ra$ and the three triplet states $|T_i(1,1)\ra$. In the (0,2) configuration only the singlet $|S(0,2)\ra$ can be populated, the triplet states $|T_i(0,2)\ra$ having much higher energies. The spin-preserving part of the Hamiltonian is given by:

\be
H_0 = \Del \ket{S_{02}}\bra{S_{02}}+ t (\ket{S_{02}}\bra{S_{11}}+\ket{S_{11}}\bra{S_{02}}).
\ee

\noi The tunneling parameter $t$ couples the two singlets allowing one electron to transfer between the two dots, while $\Del$ is the energy difference between the (0,2) - and the (1,1) -  singlets. Both parameters can be externally controlled by changing gate voltages and are used to tune the system to the desired configuration. In addition, the hyperfine interaction of the electrons with the nuclear spins mixes the singlet and triplet states in the (1,1) configuration. Each electron interacts with a large number of nuclear spins in the left (L) and right (R) dot. The global action of these nuclear spins can be included in a single operator $\hat{\textbf{B}}^N_{L,R}$ which allows us to treat the hyperfine interaction as the interaction between the electrons and this apparent magnetic field. We assume that this nuclear magnetic field remains unchanged over the typical timescale of the electron spin evolution~\cite{timenucl} and hence it can be treated as a classical magnetic field.  The combined action of the nuclear magnetic field and of an external magnetic field $B_{ext}$ is given by

\begin{widetext}
\be
H_{spin} =  B_s^z \big( \ket{T_+}\bra{T_+} - \ket{T_-}\bra{T_-} \big) + \Big(B_a^z \ket{T_0}\bra{S_{11}} + \frac{B_s^x \pm i B_s^y}{\sqrt{2}} \ket{T_0}\bra{T_{\pm}}+\frac{\mp B_a^x - i B_a^y}{\sqrt{2}}\ket{S_{11}}\bra{T_{\pm}}+\text{H.c.}\Big),
\ee
\end{widetext}

\noi where $\textbf{B}_a \equiv (\textbf{B}_L^N - \textbf{B}_R^N)/2$, $\textbf{B}_s \equiv (\textbf{B}_L^N + \textbf{B}_R^N)/2 + B_{ext} \, \textbf{z}$. The total Hamiltonian of the system is then given by $\hat{H}_{DQD} = \hat{H}_0 + \hat{H}_{spin}$~\cite{jour06}, and its dynamics can be externally controlled by tuning $\Del$, $t$ and $B_{ext}$. Assuming a large external magnetic field $B_{ext}$ the triplet states $\ket{T_i}$ are split off by the Zeeman energy, and the state space is reduced to $\left\{ \ket{T_0}, \ket{S(1,1)}, \ket{S(0,2)} \right\}$. Under these circumstances it is possible to derive an effective Hamiltonian~\cite{cois05},

\be \label{Heff}
H_{eff}=\Del \ket{S_{02}}\bra{S_{02}} + t \ket{S_{02}}\bra{S_{11}} + M \ket{S_{11}}\bra{T_0} + \text{H.c.}
\ee

\noi where $M \equiv B_a^z + \left[ (B_s^x + i B_s^y)(B_a^x - i B_a^y) + c.c \right]/ 2B_s^z $.  The eigenstates of the Hamiltonian \eqref{Heff} are given by

\be \label{veps}
 \ket{\psi_i} = \frac{\frac{M}{E_i} |T_0 \rangle + |S_{11}\rangle + \frac{t}{E_i-\Delta} |S_{02}\rangle}{\sqrt{1+\frac{M^2}{{E_i}^2}+\frac{t^2}{(E_i-\Delta)^2}}}, 
\ee 

\begin{figure} 
\scalebox{0.3}{\includegraphics{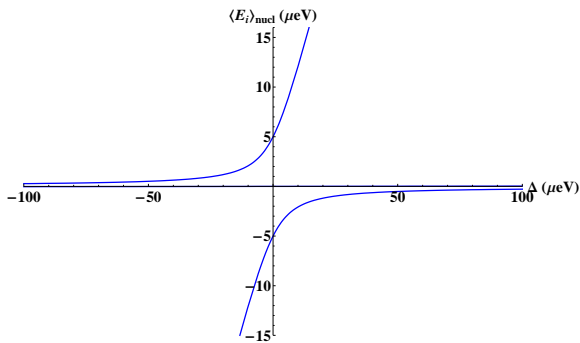}}
\caption{\label{figenergy} (Color online) Energy level diagram of the double dot system for interdot coupling $t = 5 \  \mu eV$.} 
\end{figure}

\noi where $E_i$ are their corresponding energies ($E_0 < E_1 < E_2$). These energies are given by $E_{0(1)} = \Del/3 - (s_1+s_2)/2 +(-)
i\sqrt{3} \left( s_1-s_2 \right)/2$ and $E_2 = \Del/3 + s_1 + s_2$, with $s_{1,2} = (r \pm \sqrt{q^3+r^2})^{1/3}$, $r = \Del(-18 M^2+9t^2+2\Delta^2)/54$ and $q = -(3(M^2+t^2)+\Delta^2)/9$~\cite{iroot}, and are plotted in Fig. \ref{figenergy}. 

The entangled component of the ground state is given by its projection onto the (1,1)-subspace,

\begin{equation}\label{entgs}
\ket{\psi_0^{11}} = \left( 1+M^2/E_0^2 \right)^{-1/2} \left( \frac{M}{E_0} \ket{T_0} + \ket{S_{11}} \right).
\end{equation}

\noi The amount of entanglement of $\ket{\psi_0^{11}}$ can be quantified by the concurrence~\cite{benn96}. Both the concurrence $C$ and the probability $P_{11}$ of having one electron in each dot must be large to have a highly entangled state. Their mathematical expressions are given by 

\begin{equation}\label{p11-conc}
 C=\frac{1-M^2/E_0^2}{1+M^2/E_0^2}, \ \ \  P_{11} = 1- \frac{\frac{t^2}{(E_0-\Del)^2}}{1+\frac{M^2}{{E_0}^2}+\frac{t^2}{(E_0-\Del)^2}}.
\end{equation}

\noi The concurrence $C$ is maximal when $|E_0| \gg M$, while $P_{11}$ is maximal in the presence of a large detuning ($\Del \gg |E_0|$). It can be observed in Fig. \ref{figenergy} that both conditions can only be fulfilled for intermediate values of $\Del$ where $\Del \gg |E_0| \gg M$ (typically $M \sim 0.1-1 \  \mu eV$). To find the optimal $\Del$ and $t$ values we consider the maximization of their product $\xi = P_{11} C$.  

{\it Averaging over the nuclear field distribution. -} Given the random character of the nuclear magnetic field, any observable can be calculated taking the average over the probability distribution that characterizes  $\textbf{B}^N_{L,R}$. This distribution is Gaussian with variance $\la \textbf{B}^N_{L,R} \ra = E^2_n/N_{eff}$, with $E_n \approx 0.135 \ meV$ for GaAs and $N_{eff} \approx 10^6$ for typical dots~\cite{jour06,tayl07}. For external magnetic fields $B_{ext} > 2.5 \  \mu eV \  (100 \  mT)$ the transversal components of the nuclear magnetic field ($B_{x,y}^N$) are negligible and the hyperfine interaction parameter M is mainly given by the difference of the z components of the nuclear magnetic fields ($M \simeq B_a^z$). The probability distribution for M is then approximately Gaussian with variance $\sigma^2 = \la \textbf{B}^N_{L,R} \ra /2$.

\begin{figure} 
\scalebox{0.23}{\includegraphics{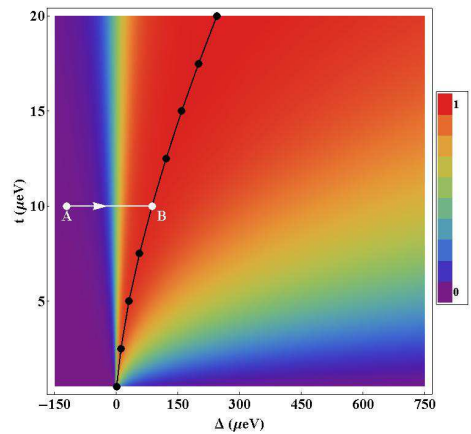}}
\caption{\label{figxi} (Color online) Averaged entanglement probability \ \ \ \ \ \ $<\xi>_{nucl}$. The black line represents the optimal configuration $\{\Del_{opt},t_{opt} \}$ that maximizes the entanglement of the system. The white solid line represents the trajectory followed from the initial state A to the final state B.} 
\end{figure}

Fig. \ref{figxi} shows the averaged success rate $\la \xi \ra_{nucl}$ obtained using the previous approximated Gaussian distribution for $M$. The entanglement of the ground state is low for $\Del<0$ where $P_{11}<0.5$,  and for small values of the interdot coupling $t$ where the ground state is nearly a separable state. The black line links the optimal values $\{ \Del_{opt}, t_{opt} \}$ that maximize $\la \xi \ra_{nucl}$. For $t_{opt}>5 \ \mu eV$, and in a large region nearby, the ground state is highly entangled ($\la \xi \ra_{nucl}>0.95$) up to values of 0.99 for $t_{opt}=20 \ \mu eV$. Obtaining the average over the nuclear probability distribution is usually a difficult task, but in this case the value of the hyperfine parameter $M$ in Eqs. \eqref{p11-conc} can to a good approximation be substituted by the standard deviation of its Gaussian probability distribution ($\la \xi(M) \ra_{nucl} \approx \xi(\sqrt{ \la M^2 \ra}) = \xi(\sigma)$). We have checked that this approximation holds in the highly entangled region and only fails for small $t$ and large $\Del$. In that region $E_0 \simeq M$, and small variations in M have a great impact on the (small) value of the concurrence $C$, and hence on $\xi$.

The entanglement of the system can be detected using witness operators. For an entangled state $\ket{\varphi}=\gamma_1 \ket{T_0} + \gamma_2 \ket{S(1,1)}$ (with $\gamma_2 > \gamma_1$) the optimal operator to detect its entanglement is given by W = \ket{T_0} \bra{T_0} + \ket{S(1,1)} \bra{S(1,1)} + \ket{T_0} \bra{S(1,1)} + \ket{S(1,1)} \bra{T_0}~\cite{guhn09}. Its expectation value is given by $Tr(W \ket{\varphi} \bra{\varphi})=(\gamma_1^2-\gamma_2^2)/2$, which is proportional to the concurrence ($Tr(W \ket{\varphi} \bra{\varphi})=-C/2$). This witness can be rewritten as $W = I/2 - \ket{S(1,1)} \bra{S(1,1)}$. In order to measure the expectation value of $W$, it is then enough to measure the probability $P[S(1,1)]$ of the ground state to be the $\ket{S(1,1)}$ state. In a double quantum dot this probability can easily be measured using a quantum point contact (QPC)~\cite{hans07}. Our proposal to detect the entanglement is then the following: To initialize the system in the $\ket{S(0,2)}$ state (point A in Fig. \ref{figxi}) and adiabatically increase $\Del$ to the optimal position $\{ \Del_{opt}, t_{opt} \}$ (point B in Fig. \ref{figxi}). A QPC can then be used to measure P(S(1,1)) and obtain the expectation value of the witness operator.

\begin{figure} 
\scalebox{0.27}{\includegraphics{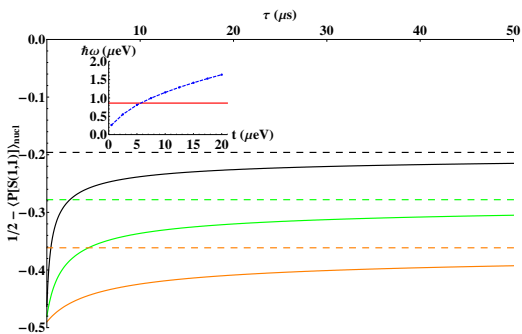}}
\caption{\label{pltwitness} (Color online) Time evolution of the expectation value of the witness operator for $T =10 \ mK$, $\alpha = 0.007$ \cite{romi07} and $t_{opt} = 20 \  \mu eV$ (orange line), $10 \  \mu eV$ (green line), $5 \  \mu eV$ (black line). Dashed lines mark their asymptotic values. Inset: Energy gap between ground and first excited states in the optimal configuration $\{\Del_{opt},t_{opt} \}$ (blue line), the red line indicates the thermal energy $k_B T = 0.86 \  \mu eV$.}
\end{figure}

{\it Time evolution of the witness operator. -} When modifying gate voltages to drive the system to its optimal position and to measure the probability $P[S(1,1)]$, the resulting charge fluctuations cause the tunnel coupling $t$ and energy offset $\Del$ to fluctuate. The inset of Fig. \ref{pltwitness} shows that $E_{gap} \geq k_B T$ for $T = 10 mK$ and $t_{opt} \geq 5 \mu eV$, (i.e. in the highly entangled region) so that quantum fluctuations are important~\cite{romi07}. In the last part of this Letter, we explore the influence of these fluctuations on the time evolution of the expectation value of the entanglement witness.

Modelling the environment as a bosonic bath~\cite{weiss99}, the Hamiltonian of the double quantum dot plus the environment is given by

\be \label{hamdec}
\hat{H}=\hat{H}_{DQD} + \hat{V}_t \hat{A}_t + \hat{V}_{\Del} \hat{A}_{\Del} + \hat{H}_{bath,t} + \hat{H}_{bath,\Del},
\ee

\noi where $\hat{V}_t = \ket{S(1,1)} \bra{S(0,2)} + \ket{S(0,2)} \bra{S(1,1)}$, $\hat{V}_{\Del}= \ket{S(0,2)} \bra{S(0,2)}$, $\hat{A}_i = \sum_k a_{i,k} (b^\dagger_{i,k} + b_{i,k})$, and $\hat{H}_{bath,i} = \sum_k  \hbar \omega_{k} (b^\dagger_{i,k} + b_{i,k})$, with $i=t,\Del$. The bosonic baths are characterized by symmetric and antisymmetric spectral functions $S^{\pm}(\omega)$ which are related by $S^+(\omega) = \coth(\hbar \omega / 2 k_b T) S^- (\omega)$. We assume a bath with Lorentzian damping $S^-(\omega)= \alpha \hbar^2 \omega \frac{1}{1+(\omega/\omega_c)^2}$, with $\omega_c$ a high cut-off frequency. We also assume weak coupling between the system and the bath, and short bath correlation times. Following the Bloch-Redfield approximation, the time-dependent reduced density matrix, written in the basis of the eigenstates of $\hat{H}_{DQD}$ is given by $\partial_t \rho_{ab} = -i \omega_{ab} \rho_{ab} + \sum_{cd} R_{abcd} \rho_{cd}$, where $\hbar \omega_{ab} = E_a - E_b$ and $R_{abcd}$ is the Bloch-Redfield tensor~\cite{weiss99}. The sum in this equation extends over terms with $\omega_{ab}-\omega_{cd} \ll 1/\Del \tau$ (the so-called secular constraint), where $\Del \tau$ is the timescale of the Markovian course-grained evolution~\cite{cohen98}. Considering the initial condition $\rho(0)=\ket{\psi_0} \bra{\psi_0}$ the only relevant components of the Bloch-Redfield tensor are $R_{0000} = -R_{1100}$ and $R_{1111} = - R_{0011}$. The time evolution for the populations of the ground and first excited states are then given by

\bea 
\rho_{00}(\tau) &=& \frac{R_{0011} + R_{1100} e^{- (R_{0011} + R_{1100}) \tau}}{R_{0011} + R_{1100}}, \label{Rcoef1}\\ 
\rho_{11}(\tau) &=& \frac{R_{1100} \left( 1 - e^{- (R_{0011} + R_{1100}) \tau} \right) }{R_{0011} + R_{1100}}. \label{Rcoef2}
\eea

\noi Rewriting the eigenstates given in \eqref{veps} as $\ket{\psi_i} = \alpha_i |T_0 \rangle + \beta_i |S_{11}\rangle + \gamma_i |S_{02}\rangle$, the coefficients $R_{iijj}$ for the optimal values $\{ \Del_{opt}, t_{opt} \}$ are approximately given by $R_{iijj} \approx 2 G(\omega_{ji}) (\beta_i \gamma_j + \beta_j \gamma_i)^2 / \hbar^2$ with $G(\omega_{ij})= 2 \pi^2 \alpha \hbar^2 \omega [1+\coth(\hbar \omega / 2 k_B T)]$. Introducing these expressions in Eq. \eqref{Rcoef1} and \eqref{Rcoef2} we can obtain the time dependent probability $P[S(1,1)](\tau) = \left| \bra{S(1,1)} \rho(\tau) \ket{S(1,1)} \right|$ which provides the expectation value of the entanglement witness plotted in Fig. \ref{pltwitness}. For these optimal values $P[S(1,1)]$  is approximately given by

\be \label{P11tau}
P[S(1,1)] \approx \rho_{00}(\tau) \alpha_1 \beta_0 \gamma_2.
\ee

\noi As can be seen in Fig. \ref{pltwitness}, the larger the value of the interdot coupling $t$ the longer the system retains a large amount of its initial entanglement. If the ground state is in its optimal configuration $\{ \Del_{opt}, t_{opt} \}$ the loss of entanglement caused by the quantum charge fluctuations in the environment becomes important for timescales of the order of microseconds~\cite{timenucl}. These times are much longer than the timescales required to manipulate electrons confined in a double quantum dot, typically of the order of 1-100 ns~\cite{hans07}. We note that for a given $t_{opt}$ the decay of the entanglement witness expectation value is smaller for $\Del < \Del_{opt}$. The loss of entanglement of the system while it is adiabatically moved to its optimal configuration is then even smaller that what is seen in Fig. \ref{pltwitness}.

{\it Conclusion. -} We have proposed a scheme to detect the entanglement between two electrons in a double quantum dot and shown that it is possible to maximize the entanglement between the two electrons in the ground state of the system. This entanglement can be detected using an entanglement witness, and its expectation value can be easily measured using a quantum point contact. We have also found that both in the optimal configuration that maximizes the entanglement and in its trajectory to this configuration, the entanglement of the ground state is robust against charge fluctuations in the environment.

We would like to thank Fabian Bodoky for useful discussions. This work was supported by the EU project MIDAS.


\begin{thebibliography}{99}
\bibitem{iroot}
We use the convention $i^{1/3}=(\sqrt{3}+i)/2$ to stablish the ordering of the eigenenergies $E_i$.
\bibitem{timenucl}
The nuclear spin configuration can be considered to be frozen for times of the order of $\mu s$ or longer (see Ref. \cite{tayl07}).
\end{thebibliography}
\end{document}